\documentstyle[12pt,A4wide,doublespace]{article}
\begin{document}

{\begin{center} 

{\bf THEORY OF CONTROLLED QUANTUM DYNAMICS} \\

\vspace{0.8cm}

Salvatore De Martino \footnote{Electronic Mail: 
demartino@vaxsa.csied.unisa.it}, 
Silvio De Siena \footnote{Electronic
Mail: desiena@vaxsa.csied.unisa.it} and
Fabrizio Illuminati \footnote{Electronic Mail: 
illuminati@vaxsa.csied.unisa.it}

\vspace{0.6cm}

{\it Dipartimento di Fisica, Universit\`{a} di Salerno, \\
     and INFN, Sezione di Napoli, Gruppo collegato di Salerno, \\
     I--84081 Baronissi (SA), Italia}

\end{center}}

\vspace{0.2cm}

{\begin{center} \large \bf Abstract \end{center}}

We introduce a general formalism, based on the stochastic
formulation of quantum mechanics, to obtain localized 
quasi--classical wave packets as dynamically controlled systems, 
for arbitrary anharmonic potentials.
The control is in general linear, and it amounts to
introduce additional quadratic and linear time--dependnt 
terms to the given potential. In this way one can construct for 
general systems either coherent packets moving
with constant dispersion, or dynamically squeezed packets
whose time--dependent dispersion remains 
bounded for all times.
In the standard operatorial framework our scheme corresponds
to a suitable generalization of the displacement and
scaling operators that generate the coherent
and squeezed states of the harmonic oscillator.

\newpage

{\bf 1. Introduction}

\vspace{0.2cm}

The present work addresses the problem of developing a
comprehensive theoretical approach to quantum control,
in the sense of providing a unified dynamical principle
to obtain the most general bounded quasi--classical 
wave--packet evolutions possible for arbitrary non harmonic
and externally driven quantum systems.

The developement of such a scheme is most easily available
and conceptually clear by working in the framework of Nelson
stochastic quantization. The latter is currently recognized as an
independent and self--consistent formulation of 
nonrelativistic quantum mechanics 
in the language of stochastic processes
\cite{nelson67}--\cite{guerra}--\cite{guerramorato}.

We remind that the experimental goal of quantum control is to 
use radiation to drive matter to a desired target or outcome
\cite{warren}, and several theoretical schemes modelling
controlled wave--packet dynamics have been suggested.
The unifying theme in all of the current theoretical and
experimental work is the optimal use of the coherence
of laser light to manipulate the quantum mechanical phase
relationship among the eigenstates of matter.

From this point of view, the coherent and squeezed states 
of the harmonic oscillator can be considered as special but
paradigmatic examples of controlled 
wave--packet dynamics \cite{garraway}. 
Progress in the femtosecond pulse technology 
allows now for the realization of quantum control in
the laboratory; for example, frequency--chirped femtosecond
laser pulses have been synthesized to control the evolution
of vibrational wave packets of the iodine molecule \cite{kohler}.

The potential interest of Nelson stochastic mechanics for the
theory of controlled wave--packet dynamics stems then from the fact
that the Nelson formulation of quantum mechanics is nothing but
a particular instance of classical stochastic control theory. Namely,
it has been proven \cite{guerramorato} that quantum dynamics can
be derived via a stochastic variational principle, by suitably
extremizing the action functional of a classical
mechanical system along diffusive trajectories
replacing the classical deterministic ones. The stochastic
variational scheme has since been extended and exploited in
a number of different contexts 
\cite{guerramarra}--\cite{morato}--\cite{loffredomorato}.

For our purposes, what is appealing of the 
stochastic quantization scheme is the possibility
that it offers to obtain both new results and 
new insights on old problems,
by looking at quantum coherence in terms of 
general properties of classical diffusion processes.

In particular, variational minimization of the stochastic
osmotic uncertainty functional yields 
the complete structure of the quantum states of
minimum uncertainty, i.e., the harmonic--oscillator coherent and
squeezed states \cite{illuminati95}. The same structure had been
derived earlier \cite{demartino94} by simply saturating the
osmotic uncertainty inequality \cite{defalco82}.

In the present paper we then address, in the stochastic quantization
approach, the problem of constructing a general
framework for controlled wave--packet dynamics, paying
special attention to the construction of coherent and squeezed
states for general non harmonic potentials.

To that purpose, we introduce the conditions of classical motion
for the wave--packet centre and the conditions of constant or
bounded time--varying dispersion as constraints for the stochastic
dynamics in an assigned, generic configurational external potential
$V(x)$. We then show that such constraints
select a class of Nelson diffusions with classical current velocity and 
wave--like propagating osmotic velocity 
\cite{demartinobis94}--\cite{demartinoter94}. 

To each Nelson diffusion belonging to such class, there is associated a
quantum state that is solution of the 
Schr\"{o}dinger equation in a
time--dependent potential $\bar{V}(x,t)$ having
a simple relation with the original system $V(x)$. In particular,
if we ask for a purely coherent wave packet solution,
i.e. whose centre $\langle \hat{x} \rangle $ follows exactly the
classical motion $x_{cl}(t)$
in $V(x)$ with constant dispersion $\Delta \hat{x}$,
then the potential $\bar{V}(x,t)$ is completely determined by the
functional form of $V(x)$ and by the knowledge of $x_{cl}(t)$.

Clearly, such ``controlling" potential $\bar{V}(x,t)$ 
may be, in principle,
experimentally fashioned in the laboratory, once a particular
potential $V(x)$ has been chosen.

Connection with the standard operatorial language is then
provided by showing that this new class of 
controlled coherent states is generated
letting the standard unitary Glauber displacement operator act
on any stationary state (for instance the
ground state) associated to $V(x)$.

We then extend our scheme to construct
coherent wave packets with bounded
time--varying dispersion $\Delta \hat{x}$. 
By resorting to
the stochastic framework we are able to determine 
the new controlling potential $\tilde{V}(x,t)$
(that in general does not coincide with the controlling
potential $\bar{V}(x,t)$ obtained in the case of constant
spreading) connected to the classical potential $V(x)$, and
to derive the evolution equation for $\Delta \hat{x}$, 
which turns out to be the classical envelope equation,
well known in the theory of charged--particle beams 
dynamics.

Knowledge of the solution $\Delta x_{cl}(t)$
of the classical envelope equation 
and of the classical trajectory $x_{cl}(t)$ associated
to the potential $V(x)$, determines unambigously the
form of the controlling potential $\tilde{V}(x,t)$.

A suitable unitary operator acting on any stationary
state (for instance the ground state)
associated to $V(x)$ is then introduced 
in the coordinate representation in order to construct
these controlled states
in the standard operatorial language: they are
displacement--operator generalized coherent states with
time--dependent dispersion.

In fact, this unitary operator acts as the product
of two distinct mappings on the ground state of the
given potential $V(x)$: the ordinary Glauber displacement 
operator and a dynamical scaling
operator, namely a dynamical squeeze operator.
Squeezing is then naturally embedded in this scheme,
and the evolution equation for $\Delta \hat{x}$ yields also the 
dynamical equation controlling the time-evolution of 
squeezing.  

Overall, the above construction provides
a natural way to introduce a class of physical coherent and 
squeezed states associated to general non harmonic potentials,
in the sense of controlled wave--packet dynamics, by imposing
suitable requirements on the physical properties of the desired
solutions. In this way we provide a specific physical
solution to the problem,
posed by Schr\"{o}dinger over seventy years ago, on how to
generalize the notion of coherent state beyond harmonic systems.

The paper is organized as follows. 
In section 2 we give a brief review of the basic ingredients of 
Nelson stochastic quantization that will be needed in the
following. In section 3 we describe the structure of the
harmonic--oscillator coherent and squeezed states in the stochastic
picture. In section 4 we show how to construct general controlled coherent
states in the stochastic framework by imposing the constrainsts
of classical motion for the wave--packet centre
and of constant wave--packet dispersion, and we derive the
explicit expression for the controlling potential
$\bar{V}(x,t)$. In section 5
we extend the discussion imposing the constraints of classical
motion and of bounded time--varying dispersion and
we derive the explicit expression for the displacement and squeeze
operators associated to the controlled coherent states
with bounded time--varying dispersion. In section
6 we draw our conclusions.

\vspace{0.6cm}

{\bf 2. Stochastic mechanics}

\vspace{0.2cm}

We shall quickly
review the basic ingredients of the stochastic formulation 
of quantum mechanics that will be needed in the following.

This quantization procedure rests on
two basic prescriptions; the first one, kinematical, promotes
the configuration of a classical system to a conservative 
diffusion process with diffusion coefficient equal 
to $\hbar/2m$.

If we denote by $q(t)$ the configurational variable for a point
particle with mass $m$, this prescription reads
\begin{equation}
dq(t) = v_{(+)}(q(t),t)dt + \sqrt{\frac{\hbar}{2m}}
dw(t) \, , \; \; \; \; dt > 0 \, \, .
\end{equation}

\noindent In the above stochastic differential equation
$v_{(+)}$ is a (forward) drift field that is determined 
by assigning the dynamics, and $w$ is the standard 
Wiener process.

An intuitive manner to look at Eq. (1) is to consider it as
the appropriate quantum form of the classical kinematical 
prescription: the Wiener process models the
quantum fluctuations that are superimposed on
the classical deterministic dynamics.

If we consider the stochastic backward increment of
the process $dq(t) = q(t) - q(t-dt)$, under
very general mathematical conditions the diffusion
$q(t)$ admits the backward representation
\begin{equation}
dq(t)=v_{(-)}(q(t),t)dt + \sqrt{\frac{\hbar}{2m}}
dw^{\ast}(t)\, ,\; \;\;\; dt>0 \, \, ,
\end{equation}

\noindent where $w^{\ast}$ is a time-reversed Wiener process.
The probability density $\rho(x,t)$ of the process,
defined on the points $x$ of
the configuration space, is induced by the conditioned
expectations on the increments of the process with respect
to the Wiener measure. The backward and forward drifts 
can then be expressed as stochastic fields on the configuration
space through the conditioned expectations
\[
v_{(+)}(x,t) \; \equiv \; \lim_{\Delta t \rightarrow 0^{+}}
\Big\langle \frac{q(t + \Delta t) - q(t)}{\Delta t}
\Big| q(t) = x \Big\rangle \, ,
\]

\begin{equation}
v_{(-)}(x,t) \; \equiv \; \lim_{\Delta t \rightarrow 0^{+}}
\Big\langle \frac{q(t) - q(t - \Delta t)}{\Delta t} 
\Big|q(t) = x \Big\rangle \, . 
\end{equation}

\noindent They represent respectively the mean forward (backward) 
velocity fields.

In the hydrodynamic picture of the process, the drifts are replaced
by the osmotic velocity $u(x,t)$,
\begin{equation}
u(x,t) \; \equiv \; \frac{v_{(+)}(x,t) - v_{(-)}(x,t)}{2} \; = \;
\frac{\hbar}{2m}\nabla [\ln\rho(x,t)] \, .
\end{equation}

\noindent and by the current velocity $v(x,t)$,
\begin{equation}
v(x,t) \; \equiv \; \frac{v_{(+)}(x,t) + v_{(-)}(x,t)}{2} \, .
\end{equation}

Finally, Fokker-Planck equation for the 
probability density $\rho(x,t)$ takes the form of the 
continuity equation
\begin{equation}
\partial_{t}\rho(x,t) \; = \; -\nabla [\rho(x,t) v(x,t)] \, .
\end{equation}

\noindent 
The process is completely determined by the 
the couple $(v(x,t),u(x,t))$, or, equivalently, by the couple
$(v(x,t), \rho(x,t))$.

The dynamical prescription is introduced by defining the 
mean regularized classical action $A$. In the hydrodynamic
Eulerian picture it is a functional of the couple $(\rho,v)$:
\begin{equation}
A\left[ \rho, v\right] \; = \; \int_{t_{a}}^{t_{b}} \left[ 
\frac{m}{2}(v^2(x,t) - u^2(x,t)) -V(x,t) \right] \rho (x,t)
d^{3}xdt \, ,
\end{equation}

\noindent where $V(x,t)$ denotes the (possibly time--dependent)
configurational external potential.

The equations of motion are then obtained by extremizing $A$
against smooth variations $\delta \rho$, $\delta v$ vanishing
at the boundaries of integration, with the continuity equation
taken as a constraint.

After standard calculations one obtains the ``quantum Newton law"
\begin{equation}
\partial_{t}v(x,t) + (v(x,t)\cdot \nabla )v(x,t) - \frac{\hbar^2}{4m^2}
\nabla\left( \frac{\nabla^2\sqrt{\rho(x,t)}}{\sqrt{\rho(x,t)}} \right) =
 -\nabla V(x,t) \, ,
\end{equation}

\noindent with the current velocity fixed to be a gradient field
at all points $x$ where $\rho(x,t) > 0$: 
\begin{equation}
v (x,t) = \frac{ \nabla S(x,t)}{m} \; ,
\end{equation}

\noindent where $S(x,t)$ is a scalar field.

Defining the wave function $\Psi(x,t)$ associated to  
a generic single--particle quantum state in the hydrodynamic form
\begin{equation}
\Psi(x,t) \; = \; 
\sqrt{\rho(x,t)}\exp \left[ \frac{i}{\hbar}S(x,t) \right] \; ,
\end{equation}

\noindent 
it immediately follows that  
the Schr\"odinger equation with potential $V(x,t)$ for the complex
wave function $\Psi(x,t)$ is equivalent to the quantum Newton law
together with the continuity equation, i.e. to two real 
nonlinearly coupled equations for the probability density $\rho(x,t)$ 
and for the phase $S(x,t)$ 
(or, alternatively, for the osmotic and current 
velocities $u(x,t)$ and $v(x,t)$).

Then, to each quantum state
there corresponds in stochastic mechanics a diffusion process $q(t)$ with
\begin{equation}
\rho(x,t) \; = \; |\Psi(x,t)|^{2} \; ,
\end{equation}

\noindent and
\begin{equation}
v(x,t) \; = \; - \frac{i\hbar}{2m}\nabla \left[ 
\ln{ \frac{\Psi(x,t)}{\Psi^{\ast}(x,t)}} \right] \, .
\end{equation}

The space integral of Eq. (8) yields the Hamilton--Jacobi--Madelung
equation. It is useful for
what follows to write this equation in the form
\begin{equation}
\partial_{t}S(x,t) + \frac{m}{2}v^{2}(x,t) - 
\frac{m}{2}u^{2}(x,t) - \frac{\hbar}{2} \nabla u(x,t) \; 
= \; - V(x,t) \; .
\end{equation}

The correspondence between expectations and correlations defined
in the stochastic and in the canonic pictures are 
\[
\langle \hat{x} \rangle = E(q(t)) \, , \; \; \; \; \; \; \; \; \; \; \;
\; \langle \hat{p} \rangle = mE(v(x,t)) \, , 
\]

\begin{equation}
\Delta \hat{x} = \Delta q \, , \; \; \; \; \; \; 
(\Delta \hat{p})^{2} = m^{2}[(\Delta u)^{2} + (\Delta v)^{2}] \, .
\end{equation}

In the above relations $\hat{x}$ and $\hat{p}$ are
the position and momentum operators in the Schr\"odinger 
picture, $\langle \cdot \rangle$ are the expectations
of the operators in the given state $\Psi$, $E(\cdot)$ is the 
expectation of the
stochastic variables in the Nelson state
$\{\rho, v\}$ corresponding to the state $\Psi$,
and $\Delta(\cdot)$ are the root mean
square deviations.

In the theory of diffusion processes, the functional 
$(\Delta q)^{2}(\Delta u)^{2}$ is known as the osmotic uncertainty
product; it shares the remarkable property that it is always greater
or equal than the square of the diffusion coefficient. The following
chain inequality then immediately follows:
\begin{equation}
(\Delta \hat{x})^{2} (\Delta \hat{p})^{2} \, \geq \,
m^{2}(\Delta q)^{2} (\Delta u)^{2} \, \geq \, 
\frac{{\hbar}^{2}}{4} \, .
\end{equation}

The osmotic 
uncertainty relation and its equivalence with the 
momentum-position uncertainty were proven in 
Ref.\cite{defalco82}. 

\vspace{0.6cm}

{\bf 3. Coherent and squeezed states of the harmonic oscillator}

\vspace{0.2cm}

Saturation of the osmotic uncertainty relation Eq. (15) yields
the Glauber coherent states in the stochastic
picture \cite{demartino94}: they are Nelson diffusions of minimum
osmotic uncertainty, with constant dispersion $\Delta q$.
They are characterized by a
deterministic classical current velocity: 
\begin{equation}
v(x,t) \, = \, \langle v(x,t) \rangle 
\, = \, v_{cl}(t) \, \equiv \, \dot{x}_{cl}(t) 
\; ,
\end{equation}

\noindent and by an osmotic velocity linear in the process:
\begin{equation}
u(x,t) \, = \, - \frac{\hbar}{2m\Delta q} \xi \, .
\end{equation}

In Eq. (16) and from now on we
denote the stochastic expectations $E(\cdot)$
with the same symbol $\langle \cdot \rangle $
used for quantum ones (wherever no confusion arises).
We have also introduced the adimensional variable 
\begin{equation}
\xi \; = \; \frac{x \, - \, x_{cl}(t)}{\Delta q} \; ,
\end{equation}
\noindent obtained by first shifting the coordinate $x$ by the classical
trajectory $x_{cl}(t)=\langle q(t) \rangle $, and then by scaling it
through the wave--packet dispersion $\Delta q$.
In fact, the adimensional configurational
variable $\xi$ will play a fundamental role in all of the following.

The probability density $\rho(x,t)$ associated to the
state (16)--(17) is readily obtained by comparing the expression for
$u(x,t)$ given in (17) and the relation (4) 
linking the probability density with 
the osmotic velocity. The phase $S(x,t)$ is obtained 
comparing the expression for $v(x,t)$ given in (16) and the relation
(9) connecting the current velocity with the phase. In this way
one reconstructs completely the wave function $\Psi(x,t)$.
The potential $V(x,t)$, with classical trajectories
$x_{cl}(t)$, that solves Schr\"{o}dinger equation
for the state (16)--(17) is finally 
identified by replacing the expressions for $S(x,t)$, $v(x,t)$,
and $u(x,t)$ in the Hamilton--Jacobi--Madelung equation (13).
This is a typical inverse--problems strategy: given a certain
state with some desired features, in this case that of being of minimum
osmotic uncertainty with constant dispersion, 
one looks for the potential that solves Schr\"{o}dinger
equation and allows for such state. 

We notice that the states of minimum osmotic uncertainty 
with constant $\Delta q$ have a purely classical current velocity,
so that $\Delta v = 0$. Therefore the minimum osmotic uncertainty
is exactly equivalent to the minimum Heisenberg uncertainty.
Then, following the procedure 
outlined above, it is straightforward to recover the complete
structure of the Heisenberg minimum uncertainty states, which
are the Schr\"{o}dinger--Glauber
coherent states of the harmonic oscillator. We have:
\[
\rho(x,t) \, = \, \frac{1}{\sqrt{2\pi(\Delta q)^{2}}}
\exp{\left[-\frac{(x - x_{cl}(t))^{2}}{2(\Delta q)^{2}}\right]} \; ,
\]
\[
S(x,t) \, = \, mv_{cl}(t)x \, - \, \frac{m}{2}v_{cl}(t)x_{cl}(t)
\, - \, \frac{1}{2}\hbar\omega t \; ,
\]
\[
\Psi(x,t) \, = \, \frac{1}{\left[ 2 \pi (\Delta q)^2
\right] ^{\frac{1}{4}}}
\exp \left\{ -{\frac{( x - x_{cl}(t))^{2}}
{4(\Delta q)^{2}}}
+{\frac{i}{\hbar}}\left[ mv_{cl}(t)\left( x-\frac{x_{cl}(t)}{2}
\right) - \frac{\hbar}{2}\omega t \right] \right\} \, ,
\]
\begin{equation}
V(x) \, = \, \frac{m}{2}\omega^{2}x^{2} \; ; \; \; \; \; \; \; \; \;
\omega^2 = \frac{\hbar^2}{4m^{2}(\Delta q)^4} \; .
\end{equation}

\noindent For those readers
unfamiliar with the stochastic picture, we remind that
in the above expressions (19) $x_{cl}(t)=\langle \hat{x}
\rangle$ and $mv_{cl}(t)=\langle \hat{p} \rangle$.

If we now impose saturation of the osmotic uncertainty, but allowing
for a time--dependent dispersion $\Delta q$, we obtain the
harmonic--oscillator squeezed states. They are Nelson
diffusions with time--varying $\Delta q$, with osmotic
velocity $u(x,t)$ still of the form (17), and with current velocity
allowing for a term dependent on the time--evolution of the dispersion
\cite{demartino94}:
\begin{equation}
v(x,t) \, = \, v_{cl}(t) \, + \, \xi \frac{d}{dt}\Delta q \, .
\end{equation}

The last term in Eq. (20) is responsible for the quantum 
anticommutator term appearing in the phase of the squeezed 
wave packets, which are quantum states of minimum Schr\"{o}dinger
uncertainty. This can be easily seen as follows: define the
centered position operator $\hat{x}_{c} = \hat{x} - \langle
\hat{x} \rangle $, the centered momentum operator $\hat{p}_{c} =
\hat{p} - \langle \hat{p} \rangle $, and the centered Nelson
process $q_{c}(t) = q(t) - \langle q(t) \rangle $. All this
quantities have zero expectation value. Next, consider the
quantity $q^{2}_{c}(t)$; obviously, $\langle q^{2}_{c}(t)
\rangle = (\Delta q)^{2}$. By straightforward calculations
one gets
\begin{equation}
\frac{d}{dt}\langle q^{2}_{c}(t) \rangle \, = \, 
\frac{d}{dt}(\Delta q)^{2} \, = \, 2\left[ \langle q(t)v(x,t)
\rangle \, - \, \langle q(t) \rangle \langle v(x,t) \rangle
\right] \, = \, \frac{ \langle \left\{ \hat{x}_{c},\hat{p}_{c}
\right\} \rangle }{m} \; ,
\end{equation}
\noindent where $\left\{ , \right\}$ denotes the quantum anticommutator,
which expresses, when taken
between $\hat{x}_{c}$ and $\hat{p}_{c}$, 
the quantum position--momentum correlation. In turn, the latter is
the Schr\"{o}dinger part of the uncertainty, and for the squeezed
states it is directly connected with the current uncertainty product
in the Nelson stochastic picture. In fact, from Eq. (20) it is
straightforward to see that 
\begin{equation}
(\Delta v)^{2} \, = \, 2\left(\frac{d}{dt}\Delta q \right)^{2} \; .
\end{equation}

The classical evolution equation for the dispersion $\Delta q$
can be easily obtained by inserting the expressions (17) and
(20) for the stochastic hydrodynamic velocities into the quantum
Newton law (8) and the Hamilton--Jacobi--Madelung equation (13);
after straightforward manipulations, we obtain
\begin{equation}
\frac{d^{2}\Delta q}{dt^{2}} \, + \, m\omega^{2}\Delta q \, =
\, \frac{\hbar^{2}}{4m^{2}(\Delta q)^{3}} \; ,
\end{equation}

\noindent which is the classical equation for the beam envelope, 
well known from the theory of classical optics and of particle
accelerator dynamics.

The expressions (19) for the Gaussian probability density and for the 
harmonic oscillator potential remain unaltered 
in the case of minimum uncertainty states with time--dependent
dispersion. However, from Eqns. (20) and (21) 
it follows that the phase of the wave function picks up an extra term
proportional to the quantum anticommutator:
\[
\Psi(x,t) \, = \, \frac{1}{\left[ 2 \pi (\Delta q)^2
\right] ^{\frac{1}{4}}}
\exp \left\{ -{\frac{( x - x_{cl}(t))^{2}}
{4(\Delta q)^{2}}} + i\left[ \frac{m}{\hbar}v_{cl}(t)x \; +
\right. \right. 
\]
\begin{equation}
\left. \left. m \left( \frac{\langle q(t)v(x,t) \rangle - 
x_{cl}(t)v_{cl}(t)}{2\hbar}\right) 
\left[ \frac{x - x_{cl}(t)}{\Delta q} \right]^{2}
- \frac{m}{2\hbar}v_{cl}(t)x_{cl}(t)
- \frac{1}{2}\omega t \right] \right\} \, .
\end{equation}

The above stochastic picture for the harmonic--oscillator
coherent and squeezed states can be entirely derived in
a stochastic variational approach by extremizing the osmotic
uncertainty product against smooth variations of the density
$\rho(x,t)$ and of the current velocity $v(x,t)$ \cite{illuminati95}.
The possibility of extending 
such variational approach to study local minimum
uncertainty behaviours in non harmonic 
systems is a current subject of investigation \cite{illuminati95}. 

This concludes our discussion of the coherent and squeezed states of
the harmonic oscillator in the framework of stochastic mechanics.
We next move to study the problem of how to generalize the concept
of coherent and squeezed states to non harmonic systems, from the
same perspective of controlled wave--packet dynamics that we have
adopted in this section, with the emphasis focused on the determination
of the potential that solves Schr\"{o}dinger dynamics once a certain
state with some desired features has been selected.

\vspace{0.6cm}

{\bf 4. Controlled coherent quantum wave packets: constant dispersion}

\vspace{0.2cm}

A well known property of the coherent states (16)--(17) and of
the squeezed states (17)--(20),
is that they follow the classical 
motion in the coherent Glauber sense:
\begin{equation}
\frac{d}{dt}(m \langle v(x,t) \rangle ) \, = \, - \nabla V(x,t)
|_{x=\langle q(t) \rangle } \; .
\end{equation}

Again, from a dynamical point of view a coherent state is a wave packet
whose center follows classical motion not only in
the mean, but even along the mean (classical) trajectories,
and whose dispersion
is either constant, or controlled in its time--evolution (squeezing).

In quantum mechanics the dynamics of mean values obeys Ehrenfest
theorem: as a consequence, the coherent evolution Eq. (25) is 
satisfied exactly if
\begin{equation}
\langle \nabla V(x,t) \rangle \;
= \; \nabla V(x,t)|_{x= \langle q \rangle} \; .
\end{equation}

In the case of quadratic potentials the above constraint
is authomatically satisfied for any quantum state.
For other generic potentials $V(x,t)$ Eq. (26) in general
cannot be satisfied. 
Our strategy is now to search, in the framework of Nelson
stochastic mechanics, for states obeying the coherence constraints
of classical motion (25)--(26) with constant dispersion $\Delta q$.

To this purpose, we first observe that Eq. (21) can be recast in
the form 
\begin{equation}
\Delta q\frac{d \Delta q}{dt} \, = \, \langle q_{c}(t) \cdot v(x,t)
\rangle \; ,
\end{equation}

\noindent where the dot denotes the scalar product between the centered
process $q_{c}(t)$ and the current velocity $v(x,t)$. 

Therefore, in
stochastic mechanics, we have that necessary and sufficient condition
for a constant dispersion $\Delta q$ is that the expectation value of
the scalar product between the centered
configurational process and the current
velocity vanishes. We see immediately that a sufficient condition for
this to happen is that the current velocity be purely classical, that is,
$v(x,t) = v_{cl}(t)$. Other possible choices of current velocities
orthogonal to the centered Nelson process could be in principle
considered, and they might lead to the definition of new classes of
states in quantum mechanics. For the moment being, we concentrate on
the simplest choice of a classical current velocity.

In stochastic mechanics, due to the existence of the continuity
equation, a particular choice of the current velocity selects an entire
class of osmotic velocities, and thus of quantum states.
In particular, the choice $v = v_{cl}(t)$ that guarantees a constant
dispersion, does not restrict the osmotic velocity to be of the
minimum uncertainty form (17). Rather, upon insertion of $v_{cl}(t)$
in the continuity equation, one can show that the latter is satisfied
by any osmotic velocity (probability density) of the following
wave--like propagating form \cite{demartinobis94}:
\begin{equation}
u(x,t) \, = \, \frac{\hbar}{2m\Delta q}G(\xi) \; ,
\end{equation}

\noindent where $G(\xi)$ can be any arbitray adimensional function 
of the adimensional shifted and scaled configurational
coordinate $\xi$, provided it yields a normalizable probability
density $\rho(\xi)$. Keeping in mind relation (4) 
connecting the density to the osmotic velocity, and reminding that
a probability density (being non negative) can be expressed as the
exponential of a real function, we have:
general form
\begin{equation}
\rho(\xi) \, = \, \frac{{\cal{N}}}{(\Delta q)^{d}}
\exp{\left[ 2R(\xi)\right] } \; .
\end{equation}

In the above, ${\cal{N}}$ is a positive, adimensional 
normalization constant,
and $d$ denotes the spatial dimension of the
system under consideration. Being 
$\nabla_{x}=\nabla_{\xi}/\Delta q$, it follows that
the adimensional normalizable
function $R(\xi)$ is related to $G(\xi)$ by 
$G(\xi)= 2\nabla_{\xi}R(\xi)$.

The associated wave function is thus of the coherent--state form
\begin{equation}
\Psi_{c}(x,t) \, = \, \frac{{\cal{N}}^{1/2}}{(\Delta q)^{d/2}}
\exp{\left[ R(\xi)+i\left( \frac{m}{\hbar}v_{cl}(t)\cdot
x + \frac{S_{0}(t)}{\hbar}\right) \right] } \; ,
\end{equation}

\noindent where we remind that 
$mv_{cl}(t)=\langle \hat{p} \rangle $ and $S_{0}(t)$ is an
arbitrary time--dependent constant.
What are the properties of this class of states
selected by the couples $(v,u)$ of the form (16)--(28)? 
By construction, they are nonstationary states of constant
dispersion. Of course, in general they are not
Heisenberg minimum uncertainty states; the
latter are recovered only with the 
choice of the linear form $G(\xi) = -\xi$. 

However, it is easily verified that
they share with the harmonic--oscillator
minimum uncertainty states the remarkable property
that they obey the constraint (25) for classical motion,
apart, at most, a constant. To see this in the simplest
and most explicit way, let us suppose that $v_{cl}(t)$ is chosen such 
that its integrals $x_{cl}(t)$ are
the classical trajectories of a generic one--dimensional
configurational potential $V(x)$. What follows can then
be generalized with minor technical complications to higher
spatial dimensions.

Suppose then that we are given 
the stationary states of the quantum mechanical
problem associated to $V(x)$. For instance, we consider
the ground state $\Psi_{0}(x,t)$. By a simple scaling
argument, it can be cast in the general form
\begin{equation}
\Psi_{0}(x,t) \, = \,
\frac{{\cal{N}}_{0}^{1/2}}{\sqrt{\Delta q_{0}}}
\exp{\left[ F\left( \frac{x}{\Delta q_{0}} \right)
+\frac{i}{\hbar}E_{0}t\right] } \; ,
\end{equation}

\noindent where ${\cal{N}}_{0}$ is the adimensional normalization
constant, $\Delta q_{0}$ the ground--state dispersion, 
$E_{0}$ the ground--state energy, and
$F$ a given adimensional function.

Comparing expressions (30) and (31) we see that the coherent
wave function $\Psi_{c}(x,t)$ is obtained from the ground state wave
function $\Psi_{0}(x,t)$ first by identifying the 
arbitrary function $R$ with $F$, and then by applying to $\Psi_{0}$
the following unitary Glauber--like displacement operator
\cite{demartinoter94}:
\begin{equation}
\hat{D}[x_{cl}(t),v_{cl}(t)] \; = \; 
\exp \left[ \frac{i}{\hbar}\left(S_{0}(t) - E_{0}t\right) \right]
\exp \left( \frac{i}{\hbar}mv_{cl}(t)\hat{x}\right) 
\exp \left( -\frac{i}{\hbar}x_{cl}(t)\hat{p} \right) \; .
\end{equation}

This operator, when applied to any
wave function $\Psi(x,t)$ displaces its space argument $x$
into $x - x_{cl}(t)$ and introduces in the phase the coherent 
term $mv_{cl}(t)x+S_{0}(t)-E_{0}t$.

We now see that the wave packet (30) is really a coherent state 
in the sense that: the associated probability density has
the same functional form of the ground--state density, so that
it shares the same statistics; in particular, the
two normalization constants coincide and the constant spreading
$\Delta q$ is just the constant ground--state dispersion 
$\Delta q_{0}$. Furthermore, the uncertainty product remains
constant too and equal to the ground--state uncertainty product,
just like the standard harmonic--oscillator coherent states do;
in fact the latter are just a particular case of the present 
construction. Last but not least, the 
wave--packet center follows the classical
motion $x_{cl}(t)$ in the given configurational
potential $V(x)$ according to the Glauber law (25).

The price to be paid for this construction is that these states do
not satisfy the time--dependent
Schr\"{o}dinger equation in the originally assigned
potential $V(x)$, unless of course the latter is chosen to be
a polynomial of degree not greater than two.

However, the controlled states $\Psi_{c}(x,t)$ of the 
coherent form (30) are still solutions
of the time--dependent Schr\"{o}dinger equation in a modified
potential $\bar{V}(x,t)$ which has a very remarkable relation to
the original potential $V(x)$. 
Namely, taking the wave--like
density $\rho(\xi)$ Eq. (29) and the coherent phase 
$S_{c}(x,t)=mv_{cl}(t)x+S_{0}(t)$ associated to the state (30),
and inserting them in the Hamilton--Jacobi--Madelung equation (13)
one finds that $\Psi_{c}(x,t)$ is a solution of the time--dependent
Schr\"{o}dinger equation in the following time--dependent
potential:
\begin{equation}
\bar{V}(x,t) \, = \, V[x-x_{cl}(t)] \, + \, m{\ddot{x}}_{cl}(t)x \; .
\end{equation}

The above expression gives the generic form
of a controlling potential that allows for 
the desired wave--packet solution with coherent
and localized dynamics in some previously
assigned configurational potential $V(x)$. 

Formally, the controlling potential $\bar{V}(x,t)$
is obtained from the original external potential $V(x)$ shifting
its argument by the classical 
trajectories associated to $V(x)$ itself,
and then by adding a correcting term linear in $x$, which is
multiplied by the inertia associated to the force field
$-\nabla_{x} V(x)$. The time--dependence of the controlling
potential is thus parametric via the 
solutions of the classical equations of motion in the
external field $V(x)$.

The question naturally arises about the physical interpretation
of this scheme of quantum control. In particular, one may ask
what is the meaning of the correcting time--dependent linear
term and more generally whether a controlling potential of
the form (33) might be fashionable in the laboratory.

In principle, the potential (33) is a well defined object, and
the practical realization of it heavily depends on the actual
choice of the original potential $V(x)$: one should then
take up a careful case--by--case analysis. However, we can 
give some general comments on the structure of $\bar{V}(x,t)$
for arbitrary $V(x)$.

We observe that the linear correcting term superimposed
on the original potential, has the simple
interpretation of an electric field with amplitude varying in time
according to the classical force law in $V(x)$. The problem
is then what actual time--dependences can be experimentally
realized.
As for the shift in the argument of $V(x)$, it is in principle
feasible, provided again
that the assigned time--dependence of the coefficients multiplying
powers of $x$ can be actually
fashioned by some wave--form generating set up; yet,
we observe that there might be many situations
where the shift could be considered
a small perturbation, allowing for Taylor expansion of $V$ or other
approximate treatements,
for instance in the case of a dynamics taking place
in spatial and/or time regions where
$x_{cl}(t)$ is a slowly varying quantity.

In conclusion, the states that we have constructed in the
stochastic framework are controlled wave packets with optimized
quasi--classical behavior, that generalize
the concept of the harmonic--oscillator coherent states in a precise
physical sense. They follow a classical motion with constant dispersion
in a given configurational potential $V(x)$ and can be obtained 
via a unitary transformation of a generic energy eigenstate of $V(x)$.
The programming potential $\bar{V}(x,t)$ that must be introduced
for the desired states to satisfy quantum dynamics has a simple
and intriguing structure, strongly related to the original potential
$V(x)$. In particular, its form allows in principle for
approximate treatements according to what degree of localization and
coherence one desires to accomplish for a generic nonharmonic
quantum system.

\vspace{0.6cm}

{\bf 5. Controlled coherent quantum wave packets: time--dependent dispersion}

\vspace{0.2cm}

We now proceed to consider the case of 
time--dependent dispersion, that is we consider the more general
form Eq. (20) for the current velocity of minimum uncertainty, and 
we follow again the strategy adopted in the previous section for
the case of the classical current velocity with constant dispersion.

We expect that the 
controlled coherent states that we will select by taking
the choice (20) for the current velocity should be
states following two coupled dynamical equations, Eq. (25) for
the wave packet center, and an evolution equation for the dispersion
$\Delta q$, analogous of the envelope equation (23) derived 
in the harmonic case.

We proceed as follows. We first set the notation by relabelling
the dispersion as a function of time. We put:
\[
\sigma(t) \; \doteq \; \Delta q \; .
\]

Next, inserting Eq. (20) into the continuity
equation (6) we are left with:
\begin{equation}
\partial_{t}\rho(x,t) \; = \; \frac{v(\xi,t)\nabla_{\xi}\rho(x,t) \,
- \, \dot{\sigma}(t)}{\sigma(t)} \; ,
\end{equation}

\noindent whose general solution is function only of $\xi$, as can
be immediately seen, e. g. by moving to Fourier space.
Again, one has selected a class of probability densities
of the form (29) as well as a class of osmotic
velocities of the wave propagating form (28).

Inserting now the current velocity (20) in the equation of 
motion (8), by Eq. (28) we obtain
\begin{equation}
-m\xi\dot{\sigma}(t)+\frac{m}{2}\nabla_{x}u^{2}(\xi) 
+ \frac{\hbar}{2}\nabla_{x}^2u(\xi) 
\;= \; \nabla_{x}V(x,t)-\langle \nabla_{x}V(x,t)\rangle \; ,
\end{equation}

\noindent where we exploited Ehrenfest theorem $\dot{v}_{cl}(t)
=-\langle \nabla_{x}V(x,t)\rangle /m$.

Letting $x=\langle q(t) \rangle = x_{cl}(t)$ (i.e. $\xi =0$),
we are left with
\begin{equation}
\nabla_{x}V(x,t)\mid_{x=x_{cl}} \, - \,
\langle \nabla_{x}V(x,t)\rangle \; = \;
\frac{m}{2}\nabla_{x}u^{2}(\xi)\mid_{\xi=0} + 
\frac{\hbar}{2}\nabla_{x}^2u(\xi)\mid_{\xi=0} \; .
\end{equation}

It is straightforward to show that this relation holds
also in the case of constant dispersion $v(x,t)=v_{cl}(t)$.
The right hand side is obviously either constant
or zero except for singular potentials: in these cases
$u(\xi)$ diverges in $\xi = 0$. However, the scheme can
be implemented also for singular potentials by taking
$\xi=x_{cl}(t)$ rather than $x=x_{cl}(t)$.
Explicit examples and applications to both singular and
non singular potentials will be discussed in detail 
elsewhere.

Given the state (20)--(28) we now want to write explicitely the 
associated phase $S(x,t)$ and the evolution equation for the 
dispersion $\sigma(t)$. This is achieved by 
exploiting Hamilton--Jacobi--Madelung equation, Eq. (9):
reminding that $v(x,t)$ is the gradient field of $S(x,t)$, Eq. (20)
implies
\begin{equation}
S(x,t) \; = \; mv_{cl}(t)x \, + \,
\frac{m[x-x_{cl}(t)]^{2}}{2\sigma(t)}\dot{\sigma}(t)
\, + \, S_{0}(t) \; .
\end{equation}

Inserting Eqs. (37), (20), and (28) in Eq. (13), 
and taking its expectation, we obtain: 
\[
\frac{m}{2}\sigma(t)\ddot{\sigma}(t)+\frac{m}{2}
\left[ {\langle v(\xi,t)\rangle }^{2}-\frac{m}{2}\langle u^{2}(\xi)
\rangle \right] \, = \, - \langle V(x,t)\rangle +x_{cl}(t)\langle
\nabla_{x}V(x,t)\rangle -\dot{S}_{0}(t) \, .
\]

\noindent We can eliminate in the above expression the 
time--derivative of the classical action and obtain, after
trivial manipulations, an evolution
equation for $\sigma(t)$:
\begin{equation}
\ddot{\sigma}(t)- \frac{\langle u(\xi)^{2}\rangle }{\sigma(t)} 
\, = \, -\frac{\langle \xi \nabla_{x}V(x,t)\rangle }{m} \, .
\end{equation}

\noindent By Eq. (28) for $u(\xi)$ it is immediately seen that
\begin{equation} 
\langle u(\xi)^2\rangle =\frac{\hbar^{2}K^2}{4m^{2}\sigma^{2}(t)} \; ,
\end{equation}

\noindent where $K^{2}=\int_{-\infty}^{\infty}G^{2}(\xi)
\rho(\xi)d\xi$; Eq. (38) is then the 
desired equation for the time-evolution
of the dispersion, generalizing the classical
envelope equation (23). Moreover, it is coupled, through the 
gradient of the potential $V(x,t)$,
with the equation of motion (25)
for the wave packet center $\langle q(t) \rangle =x_{cl}(t)$.

The general form of the wave function for this class
of states is readily obtained putting
together Eq. (29) for $\rho(\xi)$ and Eq. (37) for $S(x,t)$:
\begin{equation}
\Psi_{s}(x,t) \, = \, \frac{{\cal{N}}^{1/2}}{[\sigma(t)]^{d/2}}
\exp \left[ R(\xi) \, + \,
\frac{i}{\hbar}\left( mv_{cl}(t)x +
\frac{m[x - x_{cl}(t)]^{2}}{2\sigma(t)}\dot{\sigma}(t)
+ S_{0}(t) \right) \right] \, .
\end{equation}

We can rewrite this expression in more standard
quantum mechanical terms by reminding Eq. (21), so that
\begin{equation}
\Psi_{s}(x,t) \; = \; \frac{{\cal{N}}^{1/2}}{[\sigma(t)]^{d/2}}
\exp \left[ R(\xi) \, + \,
\frac{i}{\hbar}\left( \langle \hat{p} \rangle x +
\frac{\langle \{ \hat{q}_{c}, \hat{p}_{c} \} \rangle }
{\left( 2\sigma(t)\right) ^{2}}(x - \langle \hat{q} \rangle)^2
+ S_{0}(t) \right) \right] \, .
\end{equation}

The above wave functions $\Psi_{s}(x,t)$ are realizations
of nonstationary states with classical motion and 
controlled time--dependent spreading. They generalize
the harmonic--oscillator squeezed states (24) in the same
sense as the states (30) $\Psi_{c}(x,t)$ generalize the
harmonic--oscillator coherent states (19).

We shall now move to the study of the controlling potential 
that needs to be introduced for the 
controlled squeezed states $\Psi_{s}(x,t)$ to satisfy
Schr\"{o}dinger equation. But first
we wish to complete the analogy
with the controlled coherent states introduced in the
previous section. Namely, we will show that the states (40)
can be obtained introducing a proper scaling ``squeeze"
operator.

We proceed as follows. We first recall that the 
harmonic--oscillator squeezed states are generated by the
successive application of a scaling squeeze operator and of
Glauber displacement operator to the harmonic--oscillator 
ground state.
We then define the following dynamical scaling operator
$\hat{S}[\sigma(t)]$:
\begin{equation}
\hat{S}[\sigma(t)] \; = \; 
\exp \left[ i \left( \frac{f(t)}{\hbar} \{ \hat{q}, \hat{p} \}
\right) \right] 
\exp \left[ i \left( \frac{g(t)}{(\sigma_{0})^2}\hat{q}^{2} 
\right) \right] 
\exp \left[ \frac{f(t)g(t)}{2\hbar(\sigma_{0})^{2}}
[\{ \hat{q} , \hat{p} \} , \hat{q}^{2}] \right]   \, ,
\end{equation}

\noindent where
$\sigma_{0}$ denotes the (time--independent) dispersion
associated to the ground state $\Psi_{0}(x,t)$ of a certain
assigned configurational potential $V(x)$.
Given $\sigma(t)$ solution of Eq. (38), the
two functions $f(t)$ and $g(t)$ read
\begin{equation}
f(t) \, = \, -\frac{1}{2}\ln \left( \frac{\sigma(t)}{\sigma_{0}}
\right) \, , \; \; \; \; \; \; \; \; \; \;  g(t) \, = \, \frac{m}{\hbar}
[1-2f(t)]^{-1}\frac{d}{dt}\ln \sigma(t) \; .
\end{equation}

\noindent We see from these relations that the 
function $f(t)$ plays the role of a dynamical squeezing parameter.
We now let $\hat{S}[\sigma(t)]$ act on the ground state
wave function $\Psi_{0}(x,t)$, cast in the form (31), associated
to a given configurational potential $V(x)$. We so
define the dynamically scaled wave function
\begin{equation}
\Psi_{sc}(x,t) \; = \; 
\hat{S}[\sigma(t)]\cdot\Psi_{0}(x,t) \, .
\end{equation}

By straightforward algebra,$\{ \hat{q}, \hat{p} \} 
= i\hbar (1 + 2xd/dx)$ and 
$[\{ \hat{q},\hat{p}\} ,\hat{q}^{2}]=-4i\hbar \hat{q}^{2}$, 
and one obtains
\begin{equation}
\Psi_{sc}(x,t) \; = \; \exp \left[ f(t) \left( 1 + 
2x\frac{d}{dx} \right) \right] \cdot \chi(x,t) \, ,
\end{equation}

\noindent where $\chi(x,t)$ is given by
\begin{equation}
\chi(x,t) \; = \; 
\exp \left[ i\frac{g(t)}{(\sigma_{0})^{2}}(1-2f(t))x^{2}
\right] \Psi_{0}(x,t) \, .
\end{equation}

We now exploit the extension to the real axis of the 
following relation, holding for any analytic function 
$W(z)$, that was introduced in Ref. \cite{celeghini} 
for the study of $q$--oscillators coherent states:
\begin{equation}
Q^{x\frac{d}{dx}} \left[ W(x) \right] \; = \; W(Qx) \, ,
\end{equation}

\noindent with $Q$ a real $c$--number and $W$ 
analytic on the real axis. 
Letting $Q = \exp [2f(t)]$ and $W=\chi$
one finally is left with
\begin{equation}
\Psi_{sc}(x,t) \; = \; \exp \left[ f(t) + 
i\frac{g(t)}{({\sigma}_{0})^{2}}(1-2f(t)) e^{4f(t)}x^{2}
\right] \Psi_{0}\left( e^{2f(t)}x,t\right) \, .
\end{equation}

We then obtain the state $\Psi_{s}(x,t)$,
Eq. (40), by applying $\hat{D}[x_{cl}(t),v_{cl}(t)]$, Eq. (30), to 
$\Psi_{sc}(x,t)$:
\begin{equation}
\Psi_{s}(x,t) \; = \; \hat{D}[x_{cl}(t),v_{cl}(t)]
\left( \hat{S}[\sigma(t)]
\cdot \Psi_{0}(x,t) \right) \; = \; 
\hat{D}[x_{cl}(t),v_{cl}(t)]\cdot\Psi_{sc}(x,t) \, .
\end{equation}

By recalling Eqs. (43) it is then straightforward 
to show that $\Psi_{s}(x,t)$, Eq. (49), coincides
with the wave function (40).
We thus proved that the controlled squeezed 
states (40)--(41) can also be introduced
by a suitable modification of the
displacement--operator approach to the 
harmonic--oscillator coherent and squeezed states.

The states $\Psi_{s}(x,t)$ do not satisfy Schr\"{o}dinger
equation in the assigned potential $V(x)$, although their
wave packet center does follow the classical motion in the
force field generated by $V(x)$. 

The situation is
completely analogous to that described in the previous 
section. However now, due to the presence of a time--varying
dispersion, the controlling potential does not
coincide with the expression (33). We thus label it 
$\tilde{V}(x,t)$ to distinguish it from $\bar{V}(x,t)$
introduced in the previous section.
After solving the Hamilton--Jacobi--Madelung equation, the
new controlling potential reads:
\begin{equation}
\tilde{V}(x,t) \, = \, V[x-x_{cl}(t)] \, + \, m\left(
\ddot{x}_{cl}(t) - \frac{\ddot{\sigma}(t)}{\sigma(t)}
x_{cl}(t)\right) x \, + \, 
\frac{m\ddot{\sigma}(t)}{2\sigma(t)}x^{2} \; .
\end{equation}

We see that, compared to the controlled coherent case, a controlled
squeezed state must be associated to a programming potential with
an extra correcting quadratic term; furthermore, the time--dependent
coefficients of the correcting terms acquire a more complicated 
structure as they now depend not only on the classical trajectories 
$x_{cl}(t)$ but also on the solutions $\sigma(t)$ of the generalized
classical envelope equation (38). 

The physical interpretation of the quadratic
correcting term can be simply given in terms of a diamagnetic
interaction superimposed by the controller in the laboratory to the
previously existing external interaction $V(x)$. 
About the experimental feasibility
of such a controlling set up we can repeat in principle what we have
observed in the previous section for the linear controlling
potential. We will further comment on both the linear and the
quadratic controlling potentials in our conclusions.

\vspace{0.6cm}

{\bf 6. Conclusions and outlook}

\vspace{0.2cm}

Let us summarize at this point our results.

Working in the framework of Nelson stochastic quantization,
we have introduced a class of controlled coherent states 
(constant dispersion) and a class of controlled squeezed states
(bounded time--evolution of the dispersion).

The wave--packets centers follow a classical evolution 
in a generic preassigned time--independent
external potential $V$, while the
states obey Schr\"{o}dinger dynamics in a time--dependent
potential that has a simple relation to $V$. For constant
dispersion such time--dependent potential (that we have
named controlling or programming potential) presents a
linear correcting term to the original time--independent
potential, see Eq. (33). For the bounded time--dependent
dispersion, the controlling potential presents a linear
plus a quadratic correcting term, see Eq. (50).

For the controlled squeezed states, the evolution
equation controlling the spreading of the wave packet
is the classical envelope equation of classical optics, and it
is naturally coupled with the classical evolution equation for
the wave--packet center. The solutions of the two classical
equations enter as (time--dependent)
parameters in the programming potentials.

We have also showed that the controlled states can be obtained 
through a particular extension of
the displacement--operator coherent and squeezed states
of the harmonic oscillator by defining a suitable
dynamical scaling (squeeze) operator. 

The consequent dilatations or contractions of the 
wave--packet width are then shown to be controlled
by a single adimensional squeezing parameter $f(t)$.

A comment is due at this point.
It is well known that generalized coherent states can be obtained
extending the three different existing approaches to the definition
of the harmonic--oscillator coherent states: they are, respectively,
the minimum--uncertainty, the annihilation--operator, and the
displacement--operator method. 

The states obtained by extension of these three
methods are in general different \cite{klauderskagilmore}.

The displacement--operator coherent states
are those preserving most of the properties of
the harmonic--oscillator coherent states: they are still overcomplete
and still enjoy resolution of unity. Moreover, we have shown 
in the present paper that by a proper
choice of the Glauber parameter $\alpha$ entering the Glauber 
displacement operator, they follow 
a classical motion without dispersion.

As to squeezed states, an extension of the minimum--uncertainty
and annihilation--operator methods to arbitrary nonlinear
systems was carried out by Nieto and 
collaborators \cite{nieto93}.
They also introduced 
an extension of the minimum--uncertainty method to obtain
generalized coherent states \cite{nieto78}.

However, an extension of the displacement--operator method
to obtain generalized squeezed states runs into difficulties
\cite{katriel} and is still missing, although some progress
in that direction has been recently obtained by Nieto and
Truax for systems allowing for Holstein--Primakoff or
Bogoliubov transformations \cite{nieto95}. 

We have here introduced a possible extension of the
displacement--operator method by constructing squeezed
states via a structure of controlling potentials.
In this sense, we have given a particular answer to
the problem posed by Schr\"{o}dinger exactly 70 years ago,
whether it is possible to construct coherent--states solutions
for general non quadratic potentials. We showed that it is
possible, at the price of modifying the original potential
in a very definite way: through the external action of a
correcting interaction, the original system can follow exactly
and indefinetely a coherent (or a squeezed) dynamics.

In this paper we outlined the general features of the
method: elsewhere \cite{preparation} 
we will present the construction of
coherent and squeezed states in the sense of
Schr\"{o}dinger for a wide sample of potentials of
physical and conceptual interest.

It might be worth noting that we
have chosen for simplicity the ground state to generate
controlled coherent and squeezed states by 
the action of operators (32) and (42);
it is however immediately seen 
that their application on any stationary state yields again
controlled coherent and squeezed states of the form (30) 
and (40).

We remark that Eq. (39) represents the ``stochastic squeezing" 
condition satisfied by our states. 
Namely, it expresses the complementary 
time--dependence of the spreading $\sigma(t)$ and 
of the osmotic velocity
uncertainty $\Delta u$.

It is easily seen that in the canonic picture
Equations (14), (20), and (39) imply
$\Delta \hat q ^2 \Delta \hat p^2 = K^{2} 
+ L^{2}(t)$, with $L(t)=m
\Delta \hat q d( \Delta \hat q )/dt$. 

The reciprocal variation in time
of $\Delta \hat q$ and $\Delta \hat p$ is then ruled by 
$\Delta \hat q$ itself, determined as the solution 
of Eq. (22) with the initial condition $\Delta \hat q_{0}$.
In this way squeezing is introduced as a self-consistent
prescription on the dynamical evolution of the wave--packet 
spreading.

In conclusion, our scheme of coherent and squeezed states
via programming potentials provides an instance of the 
so--called theory of quantum control (or 
``Controlled Quantum Mechanics"), 
in the sense that, given a 
desired quantum solution (e.g. a squeezed state), one can provide
a theoretical framework to describe what dynamical system (potential)
must be introduced to produce such a state.

This theoretical model is deeply rooted in the ideas and techniques
of the theory of stochastic optimal control:
a probabilistic approach to the description and the construction
of quasi--deterministic structures, as generalized
coherent and squeezed states, must involve an optimization procedure 
(minimization of the noise) and an external dynamical
monitoring (programming potentials)
of the non--deterministic system under study (holding the
wave packet localized along the classical trajectory).

We are currently studying the possibility to apply our scheme to two
very actual and important problems in the field of quantum control:
Rydberg wave packets \cite{kosteleky}--\cite{garraway}
and molecular pseudo--gaussian states \cite{kohler}. In both
instances, the experimental situation consists of a 
laboratory--fashioned interaction (femtosecond laser pulses)
superimposed on existing natural
interactions (the atomic and molecular potentials) that gives
rise to an approximately coherent 
and localized dynamics of the wave packets. 

Our scheme of quantum control
seems potentially able to suggest from a general theoretical
framework the optimal interactions that 
should be fashioned to obtain the best
control on the dynamics (the highest degree of coherence and
localization). Namely, it would be interesting to verify the actual
experimental feasibility of the programming potentials (33)--(50)
and of the associated controlled coherent and squeezed states
for some specific physical systems such as the ones mentioned
above. We will report elsewhere \cite{preparation2} about the work
currently in progress on these subjects.

\vspace{0.6cm}

{\bf Aknowledgement}

We wish to thank Francesco Guerra for many useful
comments and suggestions.

\end{document}